# Tunable Wigner Molecules in a Germanium Quantum Dot

Chenggang Yang[1,2,3], Jun Lu[4], Hongzhang Wang[1], Jian Zeng[1,2,3], Wendong Bian[1], Zhengshan Guo[1], Jiankun Li[1], Yulei Zhang[1,5], Junwei Luo[4], and Tian Pei[1,*]

[1]*Beijing Academy of Quantum Information Sciences, Beijing 100193, China*
[2]*Beijing National Laboratory for Condensed Matter Physics, Institute of Physics, Chinese Academy of Sciences, Beijing 100190, China*
[3]*University of Chinese Academy of Sciences, Beijing 100049, China*
[4]*Institute of Semiconductor, Chinese Academy of Sciences, Beijing 100084, China*
[5]*School of Electronics, Peking University, Beijing 1000871, China*
(Dated: September 30, 2025)
*Corresponding author: peitian@baqis.ac.cn

The interplay between Coulomb interactions and kinetic energy underlies many exotic phases in condensed matter physics. In a two-dimensional electronic system, If Coulomb interaction dominates over kinetic energy, electrons condense into a crystalline phase which is referred as Wigner crystal. This ordered state manifests as Wigner molecule for few electrons at the microscopic scale. Observation of Wigner molecules has been reported in quantum dot and moiré superlattice systems. Here we demonstrate hole Wigner molecules can be formed in a gate-defined germanium quantum dot with high tunability. By varying voltages applied to the quantum dot device, we can precisely tune the hole density by either changing the hole occupancy or the quantum dot size. For densities smaller than a certain critical value, Coulomb interaction localizes individual holes into ordered lattice sites, forming a Wigner molecule. By increasing the densities, "melting" process from a Wigner molecule to Fermi liquid-like particles is observed. An intermediate configuration which indicates the coexistence of ordered structure and disordered structure can be formed within a narrow effective density range. Our results provide a new platform for further exploration of the microscopic feature of strong correlated physics and open an avenue to exploit the application of Wigner molecules for quantum information in a very promising spin qubit platform.

The transition from ordered phases to disordered phases is an attractive topic in condensed matter physics. In a two-dimensional electron system, when long range Coulomb interaction between electrons dominates over the kinetic energy of electrons, electron gas is expected to condense into a crystalline phase which is referred as Wigner crystal. Since been predicted nearly 90 years ago[1], the study of Wigner crystal has been stimulated in the last two decades by the development of materials science. Different high quality two-dimensional electron systems have been synthesized as platforms for Wigner crystal study, such as high-quality GaAs/AlGaAs quantum wells[2,3], ultra-clean bilayer graphene[4] and two-layer systems[5-7], and direct imaging in real space is realized recently[4,5,7,8]. Wigner crystal is supposed to be very fragile. By tunning the parameters, such as Wigner-Seitz radii $r_s$[7], magnetic field[9] or temperature[10], this ordered solid phase can be melted into a weakly interacting liquid phase. A lot of efforts have been made to understand this physics-rich process. However, understanding of the nature of quantum melting is still a challenging task. It is well established that direct first-order phase transition from a liquid phase to a solid phase is prohibited and intermediate phases are proposed to bridge the transition[11]. For example, electronic microemulsion phase has been reported[6], but there is still discrepancy between experimental observation and theory prediction. While mesoscopic systems more or less suffer from long range potential fluctuation, disorders etc., which hinder the clarification of the phase transition process, systems with a few electrons in microscopic scale can rule out most of those uncontrollable factors.

In the few electrons or holes regime, Wigner crystal manifests as Wigner molecule[12,13]. Electrically confined quantum dot systems have been used as a platform for Wigner molecule study. Formation and tunning of Wigner molecules have been demonstrated in quantum dot based on ultraclean carbon nanotubes[8,14,15]. Indications of Wigner molecule in quantum dot are also observed in other high-quality materials, such as InSb nanowire[16], strained GaAs[17-20], and strained Si quantum well[21]. In recent years, strained Ge quantum well has become a rising star in quantum computation due to its low disorder, strong spin-orbit interaction and small hyperfine interaction[22-25]. Tremendous progresses such as high-fidelity quantum gate above correction threshold[26], multi-qubit circuit have been realized[23,26,27]. The great potential of Ge quantum well in quantum information also boosts the optimization of material synthesis. Hole mobility over $1 \times 10^6$ cm$^2$/Vs has been achieved[28,29]. Such high-quality two-dimensional hole gas (2DHG) also makes it an ideal platform for hole-hole interaction study.

Here we demonstrate a gate-defined quantum dot formed in germanium 2DHG can accommodate tunable Wigner molecules. Depending on the effective hole density which can be precisely controlled by gate voltages, the configuration of this artificial molecule can be tuned from two-hole to three-hole. When more holes are added into the quantum dot or the size of the quantum dot shrinks, Wigner molecule is cracked and holes recover to Fermi liquid behavior. Interestingly, when quantum dot is tuned to a moderate size we observe an intermediate configuration. It contains a two-hole sub-structure which follows single-particle shell filling rules, and an unpaired hole separates spatially from it. This phenomenon is consistent with the "quantum densification" reported in a recent work[7]. Our results provide a new platform for further exploration of the microscopic feature of strong correlated physics, especially the coexistence of two competing phases, and open an avenue to exploit the application potential of Wigner molecules for quantum information in such a highly-tunable and scalable spin qubit platform.

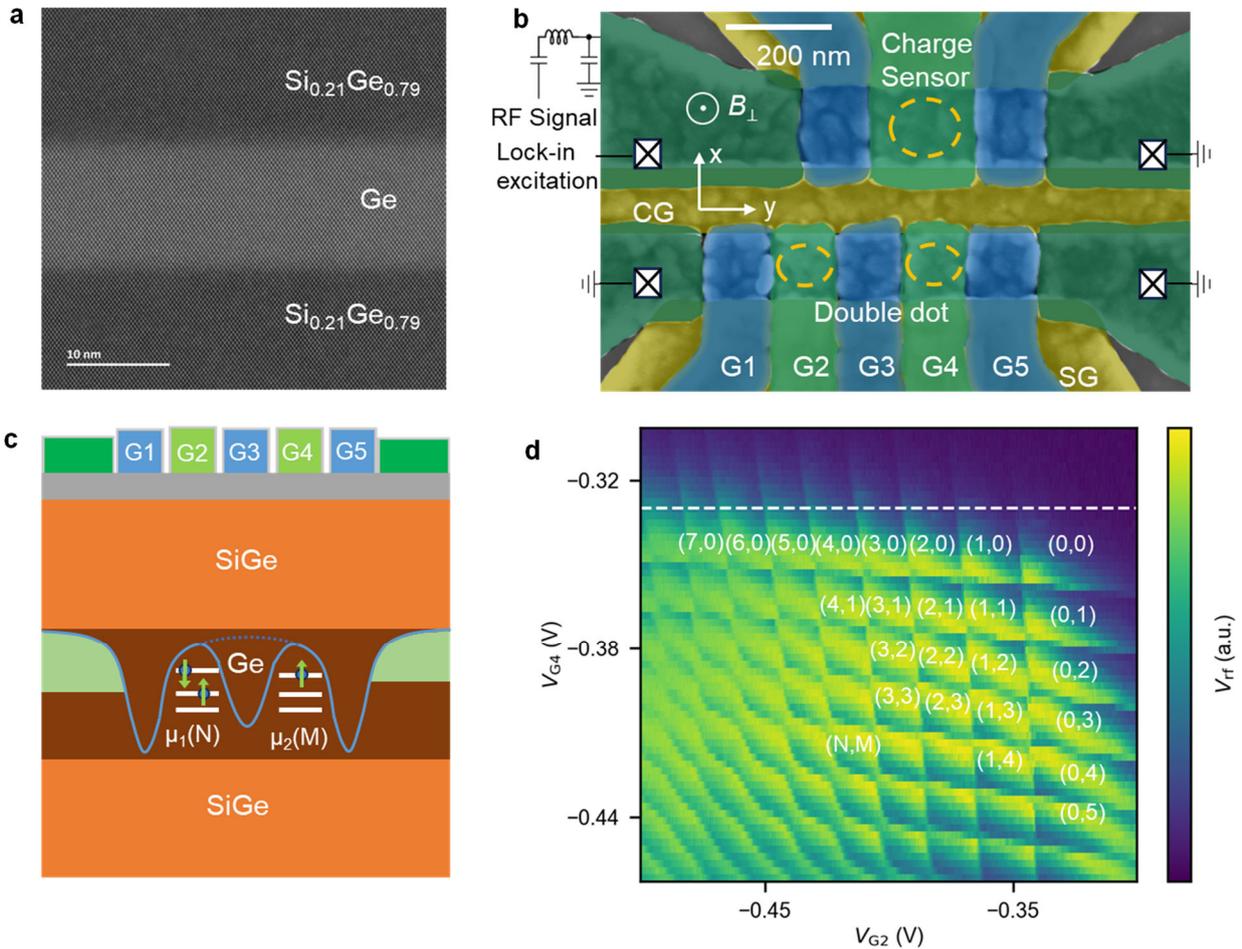

**Fig. 1. Strained germanium quantum well and gate-defined quantum dot. a**, HAADF-STEM image of the Ge/Si$_{0.21}$Ge$_{0.79}$ heterostructure grown by UHV-CVD. **b**, SEM image of a nominal identical Ge quantum dot device to that measured in the article. Yellow dashed ellipses indicate the location of quantum dots (a double dot and a nearby charge sensor used to detect the occupancy of the double dot). Four ohmic contacts are shown in the corners. The magnetic field is always applied perpendicular to the substrate (along [001]). **c,** Schematic cross section of the double dot. The potential landscape schematically shows how the double dot is defined by the confinement potential created by gates G1 to G5. Fermi sea is represented by the green filling. **d,** The charge stability diagram of the double dot. (*N*, *M*) denotes the charge occupancy of the double dot. White dashed line is chosen to study the magnetospectroscopy (Fig. 2b) of the left dot.

The germanium 2DHG is formed in the Ge/Si$_{0.21}$Ge$_{0.79}$ heterostructures grown by ultrahigh vacuum chemical vapor deposition (UHV-CVD). The aberration-corrected high-angle annular dark-field-scanning transmission electron microscopy (HAADF-STEM) image of the heterojunction is shown in Fig. 1a. The sharp interface between the quantum well and the barrier layer indicates the high quality of the 2DHG. This is furtherly proved by magnetotransport measurements (Extended Data. Fig. 1), in which hole mobility over $2 \times 10^6$ cm$^2$/Vs is achieved at a density of $\sim 2 \times 10^{11}$ cm$^{-2}$ and a signature of metal-insulator transition (MIT) is observed. The confinement along growth direction (001) and compressive strain induced by lattice mismatch between Ge quantum well and Si$_{0.21}$Ge$_{0.79}$ barrier layer lead to a large energy splitting ( $\sim$ several tens of meV) between light-hole (|3/2, ±1/2>) and heavy-hole (|3/2, ±3/2>) subband[30]. Thus, holes confined in quantum dots predominantly occupy heavy-hole states, which can be treated as a pseudospin-1/2 system (|↑> and |↓>). Standard nanofabrication process was utilized to fabricate quantum dot devices (see Method). Fig. 1b shows the false-colored scanning electron microscope (SEM) image of a device including a double dot system and a charge sensor. Two separated quantum dots (yellow dashed line above) are defined beneath the plunger gates G2 and G4 respectively. Barrier gates G1 and G5 are used to isolate quantum dots from reservoirs, while the barrier potential between two quantum dots is tuned by gate G3 (see Fig. 1c for the schematic cross section of the double dot). If the barrier height is tuned to be positive (dashed line in Fig. 1c), two dots will merge into a large single dot. The charge occupancy of two quantum dots is detected by an adjacent quantum dot charge sensor (yellow dashed line above), whose conductance changes when a hole is added to or removed from one of two quantum dots[31,32]. Either RF-reflectometry[33,34] or lock-in amplifier is used to monitor the conductance of the charge sensor. Fig. 1d shows the charge stability diagram of the double dot as a function of voltages applied to G2 and G4. Charge occupancy of the double dot is denoted as (N, M), where N (M) indicates the number of holes in the left (right) dot. The vertical (horizontal) addition lines in the stability diagram correspond to the alignment of the electrochemical potential levels of the left (right) dot ($\mu_1(N)$) ($\mu_2(M)$) shown in the schematic Fig. 1c) with the Fermi level of the reservoir.

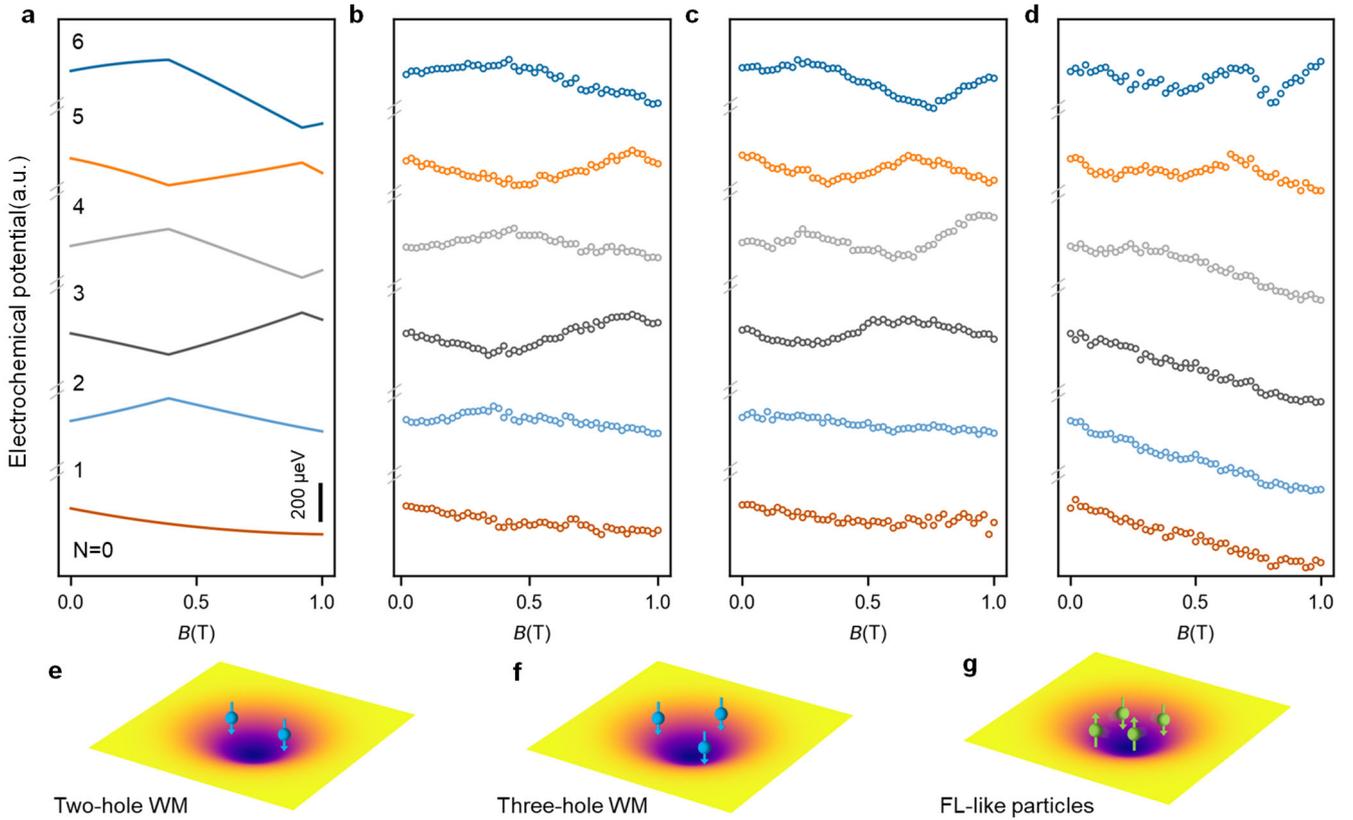

**Fig. 2. Magnetospectroscopies for first six holes in a single quantum dot with different dot size. a,** Simulation of Fock-Darwin states with an anisotropic parabolic confinement $\hbar\omega_x = 2.07$ meV and $\hbar\omega_y = 0.95$ meV. Each addition spectra is offset for clarity. **b,** $d^*$=102 nm, the quantum dot is defined under gate G2, with G4 fully depleted (white dashed line in Fig. 1d). **c,** $d^*$=133 nm, the quantum dot is also defined under G2, but larger by decreasing $V_{SG}$ and $V_{G3}$. A spin-polarized two-hole Wigner molecule (|↓↓⟩) is formed at the occupancy $N$ = 2. The addition of the third hole leads to the crossover from Wigner molecule to Fermi liquid-like particles. **d,** $d^*$ = 168 nm, G2 and G3 are combined to form a larger elliptical dot. Three-hole Wigner molecule (|↓↓↓⟩) is formed at the occupancy $N$ = 3. "Melting" occurs in the addition of the fourth hole as expected. **e, f, g,** Schematic illustration of (**e**), two-hole Wigner molecule (WM) (f), three-hole Wigner molecule (**f**) and (**g**) Fermi liquid (FL)-like particles in a quantum dot.

In this Wigner molecule study, we deplete the right dot and focus on the first six holes in the left dot (white dashed line in Fig. 1d). The size of the left dot can be estimated by disk capacitor model ($d = C/4\varepsilon_0\varepsilon_r$ see Method for detail) according to the voltage difference of addition lines. In the following article, we use effective diameter $d^* = d$ to quantify the dot size for convenience, though the real shape of the confinement potential could deform to ellipse. Here we extract $d^* = 102$ nm which is comparable with the width of the G2 electrode (120 nm, see Method). When holes are populated into the dots, they normally follow the rules from atomic physics[35], such as Pauli exclusion principle, shell filling[36], etc. The spin filling sequence was demonstrated to be well described by Fock-Darwin states. Considering an anisotropic parabolic confinement potential $V(x,y) = 1/2m^*(\omega_x^2 x^2 + \omega_y^2 y^2)$, the energy eigenvalues of the Fock-Darwin states[37]

$$E_{n_x,n_y,\uparrow(\downarrow)}(B) = \left(n_x + \frac{1}{2}\right)\hbar\omega_1 + \left(n_y + \frac{1}{2}\right)\hbar\omega_2 \pm 1/2g\mu_B B$$

where $\hbar\omega_1$ ($\hbar\omega_2$) is the perpendicular magnetic field dependent harmonic energy term in the x (y) direction, $n_x$ ($n_y$) is the corresponding quantum number (= 0, 1, 2, …), and $\pm 1/2g\mu_B B$ is the Zeeman energy for spin up (↑) (spin down (↓)). Due to the strong anisotropy of g factor for strained germanium quantum well, we apply magnetic field perpendicular to the substrate. The large g factor (~ 9.65 from Extended Data Fig. 1) enables us to easily distinguish the spin state through the slope at small magnetic field (See Extended Data Fig. 2). The magnetospectroscopy of holes in quantum dot can be characterized by measuring the shift of the addition line as a function of magnetic field, $\Delta\mu(B) = \alpha\Delta V_{G2}$ where the lever arm of G2 $\alpha = 0.115$ (see Method), and $\Delta\mu(B) = \Delta E_{n_x,n_y,\uparrow(\downarrow)}(B)$ if the confinement potential is independent of magnetic field. The spectroscopy of the first six holes in the left dot is illustrated in Fig. 2b. The field dependences of all six addition lines are well consistent with the Fock-Darwin states for anisotropic parabolic confinement potential in Fig 2a (see Extended Data Fig. 2 for detail), manifesting that holes are populated into shell orbitals with a spin sequence of $|\downarrow\rangle, |\uparrow\rangle, |\downarrow\rangle, |\uparrow\rangle, |\downarrow\rangle, |\uparrow\rangle$ at $B = 0$ T. It is worth noted that the spin sequence does not follow the Hund's first rule, where a spin filling sequence of $|\downarrow\rangle, |\uparrow\rangle, |\downarrow\rangle, |\downarrow\rangle, |\uparrow\rangle, |\uparrow\rangle$ should occur because the exchange interaction favors the spin alignment at the second shell[35,36]. This is because our quantum dot is deviated considerably from circular confinement potential ($\hbar\omega_x = \hbar\omega_y$), and thus lift the four-fold degeneracy of 2p orbital. A reasonable estimate value is $\hbar\omega_x = 2.07$ meV and $\hbar\omega_y = 0.95$ meV (See Extended Data Fig. 3), with an aspect ratio (estimated by $r = \sqrt{\frac{\hbar}{m^*\omega}}$) of the elliptical dot $r_y/r_x \sim 1.5$ (x and y-axis are defined in Fig. 1b). Hence, holes confined in the quantum dot with an effective diameter $d^* = 102$ nm follow an orbital shell filling sequence, characterized by Fermi liquid-like particles (schematic illustrated in Fig. 2g), as commonly observed[36,38].

However, an anomalous magnetospectroscopy is obtained with a larger $d^*$. By tuning the gate voltages ($V_{G3}$ and $V_{SG}$, see Extended Data Fig. 4), the shape of the barrier can be modified to enlarge the size of the quantum dot to $d^* = 133$ nm. The magnetospectroscopy of the larger dot is shown in Fig. 2c. While the spectrum of the first hole still follows Fock-Darwin model, the spectrum of the second hole shows obvious discrepancy. According to the energy shift at low magnetic field, the spin state of the second filled hole is $|\downarrow\rangle$, which is the same with the first hole. This contradicts Pauli Exclusion Principle for identical fermions occupying the same orbital, and thus implies the violation of the single-particle orbital model. However, it can be well explained by the assumption that two holes form a Wigner molecule[14]. When the confinement potential is decreased with the increase of $d^*$, the single-particle energy-level spacing $\Delta$ ($\sim d^{-2}$) becomes much smaller than the Coulomb repulsion energy $U$ between two holes. Two holes are thus separated from each other (as shown in Fig. 2e), forming a spin-polarized Wigner molecule with a total spin $S = 1$. The spectrums of the third hole to the sixth hole in Fig. 2c are similar to those in Fig. 2a, with the kinks moving to lower magnetic field because of the reduced harmonic potential energy. This means when one more hole is added, the increased effective hole density leads to the crack of the two-hole Wigner molecule, and all holes recover to Fermi liquid-like particles. A microscopic "melting" phenomenon is thus observed in a quantum dot, analogous to the mesoscopic melting of Wigner crystal to Fermi liquid. We note that at small magnetic field (0 ~ 0.24 T) the spectrum of the fourth hole is not consistent with that of Fig. 2a. Such negative-positive slope reversal before the first avoid-cross is a feature of the fourth hole in a circular quantum dot (See Extended Data Fig. 2). This is reasonable as the confinement along x direction is considerably reduced ($V_{SG}$ = 0.15 V, 300 mV lower than that in Fig. 2b, see Extended Data Fig. 4). And thus the shape of confinement is circular for first four holes, then transform to elliptical for more holes[36,39] (such as the fifth and the sixth).

To observe Wigner molecule consists of more holes, a natural way is to decrease the hole density at a given occupancy (such as $N = 3$) in quantum dot, which can be also realized by increasing $d^*$. This is realized by combining G2 and G3 as one plunger gate to form a larger elliptical quantum dot. The charge stability diagram is shown in Fig. 3a. The effective diameter $d^*$ can be changed by adjusting $V_{G4}$. We measure the magnetospectroscopy of the quantum dot whose effective diameter $d^*$ = 168 nm, as shown in Fig. 2d. The magnetic field dependences of the first three holes are nearly identical, suggesting a three-hole Wigner molecule with $S = 3/2$ is formed (schematically illustrated in Fig. 2f). A three-hole Wigner molecule is normally triangular configuration, which is directly imaged in moiré superlattice[13]. Holes recover to Fermi-liquid behavior with the addition of the fourth hole (larger hole density), as expected.

We now turn to the intermediate effective dot size where the competition between the Coulomb interaction and kinetic energy is pronounced in the three holes regime. We set $V_{G4}$ = -0.105 and -0.175 V (orange and blue dashed line in Fig. 3a) to adjust $d^*$ = 136 nm and 147 nm, respectively. Corresponding magnetospectroscopies are shown in Fig. 3b and Fig. 3c. The spectrums of the first, second, and fourth holes exhibit similar magnetic field dependence, whereas the third addition line shows a subtle distinction. In Fig. 3b, the third addition line displays a negative-positive-negative slope reversal, consistent with the trend of the third addition line modeled with Fock-Darwin state in Fig. 2a. However, Fig. 3c shows a positive-negative slope reversal (red dashed line) for the third addition line, which is in accordance with the magnetic field dependence of the second addition line in Fig. 2a. This implies two very different physics pictures. For quantum dot with $d^*$ = 136 nm, spin polarized two-hole Wigner molecule is formed when two holes are added. The addition of the third hole makes two-hole Wigner molecule "melt" and all three holes agglomerate to transit to liquid-like state, similar to the case in Fig. 2c. For slightly larger quantum dot with $d^*$ = 147 nm, the addition of the third hole fails to break the two-hole Wigner molecule, nor establishes a three-hole Wigner molecule. In this case, two of the three holes form a liquid droplet, occupying the same single-particle orbital with opposite spin, while an unpaired hole still separates spatially from them. As the number of holes furtherly increase to four, the kinetic energy starts dominating and all holes again recover to Fermi-liquid behavior. The intermediate three-hole configuration is well consistent with the "quantum densification" reported in a recent work[7]. In their disordered Wigner solid[7], the authors have imaged the coalescence of hole and partially delocalized hole wave functions in local regions during the transition between two well-defined Wigner solid lattice configurations. Our result verifies the quantum nature of these partially delocalized holes in a microscopic picture.

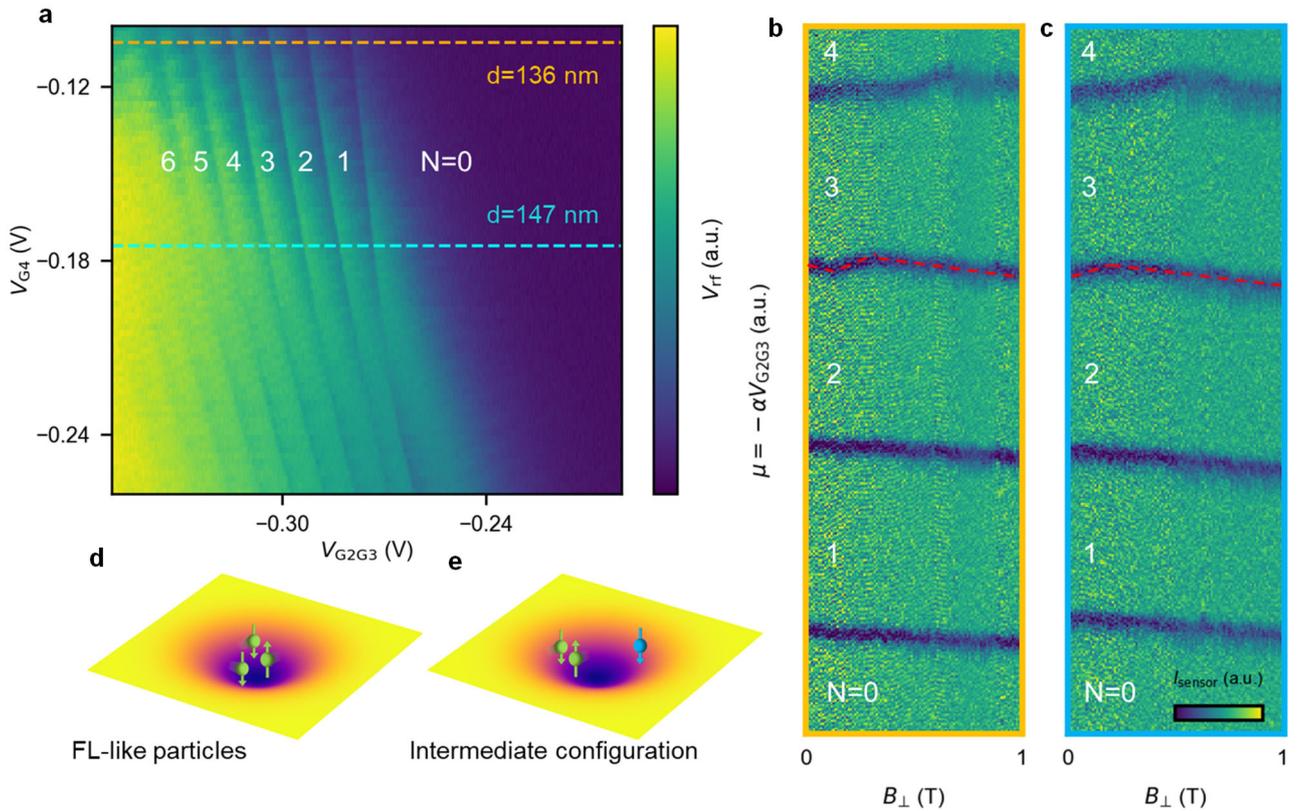

**Fig. 3. Observation of an intermediate configuration in a transition regime. a,** Charge stability diagram of an elliptical dot combined with G2 and G3. $d^*$ is changed with the gate voltage of G4. **b, c,** Magnetospectroscopy for the first four holes in quantum dot with $d^*$=136 nm (**b**) and $d^*$=147 nm (**c**). The subtle distinction in the third addition line is emphasized with red dashed line. An intermediate configuration is formed at the hole occupancy $N = 3$ for $d^*$ = 147 nm. **d, e,** Schematic illustration of Fermi liquid-particles (**d**), intermediate configuration (**e**) for $N = 3$.

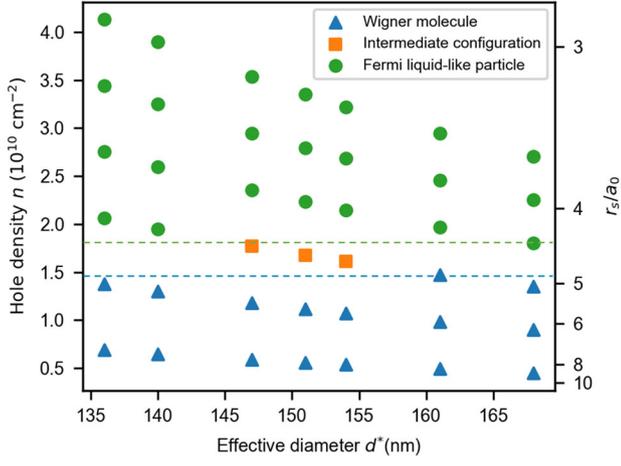

**Fig. 4. Phase diagram of Wigner molecules.** For every $d^*$, each point corresponds to the number of holes (from $N$ = 1 to 6) and is classified as Wigner molecule (blue triangles), intermediate configuration (orange squares) or Fermi liquid-like particles (green circles). Green and blue dashed line represents the critical density of Wigner molecule and Fermi liquid-like particles, respectively.

We summarize the density dependence of Wigner molecules in a phase diagram displayed in Fig. 4. The effective density $n$ is defined as $n = \frac{N}{\pi d^{*2}/4}$, determining the Wigner-Seitz radii $r_s = \frac{m^* e^2}{4\pi\varepsilon\hbar^2\sqrt{\pi n}}$, where $m^*$ is the hole effective mass, $\varepsilon$ is the effective dielectric constant. For every effective diameter $d^*$, each data point corresponds to the number of holes (from $N$ = 1 to 6) and is classified as Wigner molecule, intermediate configuration or Fermi liquid-like particles according to their magnetospectroscopies. The crossover from Wigner molecule to Fermi liquid shows dependence on effective density $n$ instead of the number of holes $N$. This density-driven transition verifies the validity of our analysis for spin-polarized Wigner molecule from magnetospectroscopy and is a sign of the very low disorder in our quantum dot. The critical density $n_W$ of the Wigner molecule in our germanium quantum dot is $1.47 \times 10^{10}$ cm$^{-2}$, which corresponds to $r_{s,W}$ = 4.85. While this value is an order of magnitude smaller than the calculated $r_{s,2D} \approx 38$ for an ideal Wigner crystal in two-dimensional electronic systems[40], it agrees well with the Monte Carlo simulation-based prediction for crossover from Wigner molecular to Fermi liquid in parabolic quantum dots, where $r_{s,W} \approx 4$ is yielded[41]. Our intermediate configuration appears from $n$ = $1.61 \times 10^{10}$ cm$^{-2}$ to $n$ = $1.77 \times 10^{10}$ cm$^{-2}$. Thus the relative density width $\Delta n/n$ is about 0.1. This is similar to the range for the solid-liquid mixture phase observed in bilayer MoSe$_2$, but also much larger than that for the microemulsion phase in an ideal two-dimensional electronic system[42]. A new theoretical model that takes disorders and confinement into account is thus required to explain this discrepancy.

In conclusion, we report a quantum dot confined in a very low disorder germanium quantum well can accommodate tunable Wigner molecules. The strong hole-hole interaction and strong anisotropy of g factor allow us to realize and detect spin-polarized Wigner molecules through magnetospectroscopies. We achieve the transition from Fermi liquid-like particles to spin-polarized Wigner molecule by tuning effective hole densities in the quantum dot. We also observe a three-hole intermediate configuration in the regime where the competition between the Coulomb interaction and kinetic energy is pronounced. All these crossovers are density-driven which can be precisely controlled by gate voltages (i.e., increasing the hole number or tuning the confinement potential). This versatility provides an excellent platform for further exploration of the microscopic feature of strong correlated physics, such as the controversial microemulsion phase, the simulation of Wigner molecule-relevant exotic phases[18,43], etc. Understanding the Wigner molecule in germanium quantum dot is also significant for the development of scalable quantum computer, for the Pauli spin blockade[44], the key ingredient of qubit readout scheme[32,45] can be hampered by Wigner molecularization. Fortunately, it can be addressed by adding an even number of holes to recover Fermi liquid-particles, such as from (1, 1) → (2, 0), to (3, 1) → (4, 0). Thus, it is possible to integrate the potential of Wigner molecules into the present mature qubit platform, such as higher-level dynamic nuclear polarization[20], an intriguing method to enhance coherence time[46,47].

## Methods

**Heterostructure growth**

Different with previously reported strained germanium quantum wells (grown by reduced pressure chemical vapor deposition (RP-CVD)[29,48,49], low-energy plasma chemical vapor deposition (LEPE-CVD)[24], and molecular beam epitaxy (MBE)[50]), our heterostructures are grown by ultrahigh-vacuum chemical vapor deposition (UHV-CVD). The heterostructure was fabricated starting with a two-step Ge buffer, followed by a thick, relaxed graded SiGe buffer and a uniform SiGe buffer grown at 675 °C. Subsequently, a 150 nm $Si_{0.21}Ge_{0.79}$ bottom barrier, a 13 nm strained Ge quantum well, a 65 nm- $Si_{0.21}Ge_{0.79}$ top barrier, and finally a 1-2 nm Si cap were deposited at 450 °C (see details in the previous work[51]).

**Device Fabrication**

A mesa is first defined on a substrate by reactive ion etching to isolate the active areas and suppress leakage currents. Ohmic contacts are fabricated by first patterning S1813 photoresist with a laser direct writing lithography system, followed by a selective removal of the Si cap using buffered oxide etch (BOE). A 40-nm-thick Al layer was deposited via electron-beam evaporation, followed by lift-off, and annealed at 320 °C for 2 hours.

For quantum dot devices, a 20 nm-thick gate dielectric is grown by atomic layer deposition. We use electron-beam lithography and electron-beam evaporation to fabricate three layers of overlapping aluminum gates, each with a native oxidation to form inter-gate dielectrics. The designed gate size for G1 to G5 are all 150×120 nm (150 nm for x-axis and 120 nm for y-axis).

For Hall-bar devices, we grow a 35 nm-thick gate dielectric, and deposit a top gate with a 20 nm-Ti 160 nm-Au.

**Experimental Set-up**

Magnetotransport experiments for Hall-bar devices is performed in a Physical Property Measurement System (PPMS, Quantum Design) at 1.7 K. All measurements for quantum dot device are performed in a Bluefors dilution refrigerator with a base temperature of 13 mK. The d.c. voltages on gates are provided by a Qdevil QDAC high resolution digital-to-analogue converter. The current through charge sensor is detected by Stanford Research Systems SR865A lock-in amplifier with an excitation of 50 μV at a 7.5 Hz reference frequency. RF signal provided by a Zurich instruments UHFLI is applied to the accumulation gate of charge sensor to reduce leakage[52]. The reflected signal is amplified with LNF Cryogenic Low Noise Amplifier at 4K-cold plate and then demodulated by UHFLI. An arbitrary waveform generator (AWG) Tektronix AWG5208 supplies a square waveform to gate G2 through a bias tee on PCB, for pulse-gate experiment in Extended Fig. 3. A perpendicular magnetic field is applied through an American Magnetics three-axis magnet.

**Magnetospectroscopy Acquisition**

We detect the current of charge sensor through lock-in amplifier for magnetospectroscopy measurements. The magnetospectroscopy in Fig. 2 is obtained by stepping magnetic field and sweeping voltage in a fixed compensation on charge sensor[53]. The data points are acquired by fitting voltage scan for each magnetic field to[54]

$$\frac{dI_{sensor}}{dV} = a + b[\cosh(\frac{V - V_{peak}}{c})]^{-2}$$

where a, b and c are fitting parameters, and $V_{peak}$ is defined as the location of an addition line, contributing to each data point. It is worth noted that the Coulomb peaks of charge sensor are strongly modulated at some magnetic field values, which may relate to the Landau levels. Consequently, the data points are not smoothly changed at such magnetic field values. For a better comparison, the magnetospectroscopy in Fig. 3 is obtained using a feedback loop compensation[53] on charge sensor to keep the sensitivity maximum.

**Dot size Estimation**

The size of quantum dot is estimated by disk capacitor model, $C = 4\varepsilon_0\varepsilon_r d$, where $C$ is the total capacitance of the quantum dot, $\varepsilon_0 = 8.85 \times 10^{-12}$ F/m, $\varepsilon_r = 16.3$ for the relative dielectric constant of germanium, and $d$ is the diameter of a circular quantum dot. As the article mentioned, we use effective diameter $d^* = d$ to quantify the dot size independent of the real shape of dot. In the constant interaction model[32], the charging energy $E_C = e^2/C$ and the addition energy $E_{add}(N) = \mu(N+1) - \mu(N) = E_C + \Delta E$, where $\Delta E$ is energy spacing between two discrete quantum levels. For $N = 1$, $\Delta E = 0$ and $E_{add}(1) = E_C$ due to two-fold spin degeneracy at ground state. Thus, we can estimate $d$ with the electrochemical potential difference (determined by scaling the voltage difference $\Delta V$ with a lever arm $\alpha$) between two addition lines ($N = 2$ and $N = 1$) at $B = 0$. The lever arm $\alpha$ is extracted as 0.115 for $V_{G2}$ with a d.c. bias applied to the source of double dot (see Extended Data Fig. 5), and 0.18 for $V_{G2G3}$ determined by the broaden of addition line[36]. However, the constant interaction model is invalid when Wigner molecule is formed. Instead, we use $E_{add}(N_m-1)$ to calculate $d$, where $N_m$ is the hole occupancy at which holes just melt to Fermi liquid-like particles. For example, the effective diameter $d^*$ of Fig. 2c and Fig. 3b are estimated by $E_{add}(2)$,

Fig. 2d and Fig. 3c by $E_{add}(3)$, etc. In fact, for a quantum dot that can host Wigner molecule, the single particle energy level ($\sim \Delta E$) is much smaller than Coulomb interaction ($\sim E_C$), and therefore $E_{add}(N_m-1)$ should approximate $E_C$.

## Acknowledgements

We Thank the useful discussion with Wenkai Lou and Kai Chang from Institute of Semiconductor, Chinese Academy of Sciences. This work was supported by the National Natural Science Foundation (Grant No. 12374469) and the National Key Research and Development Program (Grant No. 2024YFA1408200).

## Author Contributions

C. Yang performed the experiments and analysed the data with the help from T. Pei. J. Lu grew the Ge/SiGe heterostructure. H. Wang and W. Bian fabricated the devices. J. Zeng and Z. Guo helped with the measurement system setup. J. Li and Y. Zhang characterised the magnetotransport experiments of the 2DHGs. J. Luo reviewed and revised the article. T. Pei conceived and supervised the project. T. Pei and C. Yang wrote the paper with input from all authors.

## Competing interests

The authors declare no competing interests.

## Reference

1. Wigner, E. On the Interaction of Electrons in Metals. *Physical Review* **46**, 1002-1011, doi:10.1103/PhysRev.46.1002 (1934).
2. Zhao, L. *et al.* Dynamic Response of Wigner Crystals. *Physical Review Letters* **130**, 246401, doi:10.1103/PhysRevLett.130.246401 (2023).
3. Goldman, V. J., Santos, M., Shayegan, M. & Cunningham, J. E. Evidence for Two-Dimensional Quantum Wigner Crystal. *Physical Review Letters* **65**, 17 (1990).
4. Tsui, Y.-C. *et al.* Direct observation of a magnetic-field-induced Wigner crystal. *Nature* **628**, 287-292, doi:10.1038/s41586-024-07212-7 (2024).
5. Li, H. *et al.* Imaging two-dimensional generalized Wigner crystals. *Nature* **597**, 650-654, doi:10.1038/s41586-021-03874-9 (2021).
6. Sung, J. *et al.* An electronic microemulsion phase emerging from a quantum crystal-to-liquid transition. *Nature Physics* **21**, 437-443, doi:10.1038/s41567-024-02759-8 (2025).
7. Xiang, Z. *et al.* Imaging quantum melting in a disordered 2D Wigner solid. *Science* **388**, 736 (2025).
8. Shapir, I. *et al.* Imaging the electronic Wigner crystal in one dimension. *Science* **364**, 870 (2019).
9. Chen, Y. P. *et al.* Melting of a 2D quantum electron solid in high magnetic field. *Nature Physics* **2**, 452-455, doi:10.1038/nphys322 (2006).
10. Zong, Q. J. *et al.* Quantum melting of generalized electron crystal in twisted bilayer MoSe2. *Nature Communications* **16**, 4058, doi:10.1038/s41467-025-59365-2 (2025).
11. Spivak, B. & Kivelson, S. A. Phases intermediate between a two-dimensional electron liquid and Wigner crystal. *Physical Review B* **70**, doi:10.1103/PhysRevB.70.155114 (2004).
12. Nazarov, Y. V. & Khaetskii, A. V. Wigner molecule on the top of a quantum dot. *Physical Review B* **49**, 5077, doi:10.1103/PhysRevB.49.5077 (1994).
13. Li, H. *et al.* Wigner molecular crystals from multielectron moiré artificial atoms. *Science* **385**, 86 (2024).
14. Deshpande, V. V. & Bockrath, M. The one-dimensional Wigner crystal in carbon nanotubes. *Nature Physics* **4**, 314-318, doi:10.1038/nphys895 (2008).
15. Pecker, S. *et al.* Observation and spectroscopy of a two-electron Wigner molecule in an ultraclean carbon nanotube. *Nature Physics* **9**, 576-581, doi:10.1038/nphys2692 (2013).
16. Kristinsdóttir, L. H. *et al.* Signatures of Wigner localization in epitaxially grown nanowires. *Physical Review B* **83**, 041101(R), doi:10.1103/PhysRevB.83.041101 (2011).
17. Stewart, D. R., Sprinzak, D., Marcus, C. M., Duruöz, C. I. & S., H. J. J. Correlations Between Ground and Excited State Spectra of a Quantum Dot. *Science* **278**, 1784 (1997).
18. Ho, S.-C. *et al.* Imaging the Zigzag Wigner Crystal in Confinement-Tunable Quantum Wires. *Physical Review Letters* **121**, 106801, doi:10.1103/PhysRevLett.121.106801 (2018).
19. Yamamoto, M., Stopa, M., Tokura, Y., Hirayama, Y. & Tarucha, S. Negative Coulomb Drag in a One-Dimensional Wire. *Science* **313**, 204 (2006).
20. Jang, W. *et al.* Wigner-molecularization-enabled dynamic nuclear polarization. *Nature Communications* **14**, 2948, doi:10.1038/s41467-023-38649-5 (2023).
21. Corrigan, J. *et al.* Coherent Control and Spectroscopy

22 Hendrickx, N. W., Franke, D. P., Sammak, A., Scappucci, G. & Veldhorst, M. Fast two-qubit logic with holes in germanium. *Nature* **577**, 487, doi:10.1038/s41586-019-1919-3 (2020).

23 Hendrickx, N. W. *et al.* A four-qubit germanium quantum processor. *Nature* **591**, 580, doi:10.1038/s41586-021-03332-6 (2021).

24 Jirovec, D. *et al.* A singlet-triplet hole spin qubit in planar Ge. *Nature Materials* **20**, 1106-1112, doi:10.1038/s41563-021-01022-2 (2021).

25 Hendrickx, N. W. *et al.* Sweet-spot operation of a germanium hole spin qubit with highly anisotropic noise sensitivity. *Nature Materials* **23**, 920-927, doi:10.1038/s41563-024-01857-5 (2024).

26 Wang, C.-A. *et al.* Operating semiconductor quantum processors with hopping spins. *Science* **385**, 447 (2024).

27 Zhang, X. *et al.* Universal control of four singlet–triplet qubits. *Nature Nanotechnology* **20**, 209-215, doi:10.1038/s41565-024-01817-9 (2024).

28 Stehouwer, L. E. A. *et al.* Germanium wafers for strained quantum wells with low disorder. *Applied Physics Letters* **123**, 092101, doi:10.1063/5.0158262 (2023).

29 Kong, Z. *et al.* Undoped Strained Ge Quantum Well with Ultrahigh Mobility of Two Million. *ACS Appl Mater Interfaces* **15**, 28799-28805, doi:10.1021/acsami.3c03294 (2023).

30 Hendrickx, N. W. *Qubit arrays in germanium.* PhD thesis, Delf University of Technology, (2021).

31 Elzerman, J. M. *et al.* Few-electron quantum dot circuit with integrated charge read out. *Physical Review B* **67**, 161308(R), doi:10.1103/PhysRevB.67.161308 (2003).

32 Hanson, R., Kouwenhoven, L. P., Petta, J. R., Tarucha, S. & Vandersypen, L. M. K. Spins in few-electron quantum dots. *Reviews of Modern Physics* **79**, 1217-1265, doi:10.1103/RevModPhys.79.1217 (2007).

33 Schoelkopf, R. J., Wahlgren, P., Kozhevnikov, A. A., Delsing, P. & Prober, D. E. The Radio-Frequency Single-Electron Transistor(RF-SET): A Fast and Ultrasensitive Electrometer. *Science* **280**, 1238 (1998).

34 Vigneau, F. *et al.* Probing quantum devices with radio-frequency reflectometry. *Applied Physics Reviews* **10**, 021305, doi:10.1063/5.0088229 (2023).

35 Kouwenhoven, L. P., Austing, D. G. & Tarucha, S. Few-electron quantum dots. *Reports on Progress in Physics* **64**, 701 (2001).

36 Liles, S. D. *et al.* Spin and orbital structure of the first six holes in a silicon metal-oxide-semiconductor quantum dot. *Nature Communications* **9**, 3255, doi:10.1038/s41467-018-05700-9 (2018).

37 Madhav, A. V. & Chakraborty, T. Electronic properties of anisotropic quantum dots in a magnetic field. *Physical Review B* **49**, 8163-8168, doi:10.1103/PhysRevB.49.8163 (1994).

38 Tarucha, S., Austing, D. G., Honda, T., Hage, R. J. v. d. & Kouwenhoven, L. P. Shell Filling and Spin Effects in a Few Electron Quantum Dot. *Physical Review Letters* **77**, 3613 (1996).

39 van Riggelen, F. *et al.* A two-dimensional array of single-hole quantum dots. *Applied Physics Letters* **118**, 044002, doi:10.1063/5.0037330 (2021).

40 Drummond, N. D. & Needs, R. J. Phase Diagram of the Low-Density Two-Dimensional Homogeneous Electron Gas. *Physical Review Letters* **102**, 126402, doi:10.1103/PhysRevLett.102.126402 (2009).

41 Egger, R., Häusler, W., Mak, C. H. & Grabert, H. Crossover from Fermi Liquid to Wigner Molecule Behavior in Quantum Dots. *Physical Review Letters* **82**, 16 (1999).

42 Joy, S. & Skinner, B. Upper bound on the window of density occupied by microemulsion phases in two-dimensional electron systems. *Physical Review B* **108**, L241110, doi:10.1103/PhysRevB.108.L241110 (2023).

43 Goldberg, A., Yannouleas, C. & Landman, U. Electronic Wigner-molecule polymeric chains in elongated silicon quantum dots and finite-length quantum wires. *Physical Review Applied* **21**, 064063, doi:10.1103/PhysRevApplied.21.064063 (2024).

44 Ono, K., Austing, D. G., Tokura, Y. & Tarucha, S. Current Rectification by Pauli Exclusion in a Weakly Coupled Double Quantum Dot System. *Science* **297**, 1313 (2002).

45 Burkard, G., Ladd, T. D., Pan, A., Nichol, J. M. & Petta, J. R. Semiconductor spin qubits. *Reviews of Modern Physics* **95**, 025003, doi:10.1103/RevModPhys.95.025003 (2023).

46 Foletti, S., Bluhm, H., Mahalu, D., Umansky, V. & Yacoby, A. Universal quantum control of two-electron spin quantum bits using dynamic nuclear polarization. *Nature Physics* **5**, 903-908, doi:10.1038/nphys1424 (2009).

47 Cai, X., Waleligin, H. Y. & Nichol, J. M. The formation

of a nuclear-spin dark state in silicon. *Nature Physics* **21**, 536-541, doi:10.1038/s41567-024-02773-w (2025).

48  Sammak, A. *et al.* Shallow and Undoped Germanium Quantum Wells: A Playground for Spin and Hybrid Quantum Technology. *Advanced Functional Materials* **29**, 1807613, doi:10.1002/adfm.201807613 (2019).

49  Stehouwer, L. E. A. *et al.* Exploiting strained epitaxial germanium for scaling low-noise spin qubits at the micrometre scale. *Nature Materials*, doi:10.1038/s41563-025-02276-w (2025).

50  Zhang, J. *et al.* High-quality Ge/SiGe heterostructure with atomically sharp interface grown by molecular beam epitaxy. *Applied Physics Letters* **125**, 122106, doi:10.1063/5.0210639 (2024).

51  Zhang, D. *et al.* Sharp interface of undoped Ge/SiGe quantum well grown by ultrahigh vacuum chemical vapor deposition. *Applied Physics Letters* **121**, 022102, doi:10.1063/5.0097846 (2022).

52  Liu, Y. Y. *et al.* Radio-Frequency Reflectometry in Silicon-Based Quantum Dots. *Physical Review Applied* **16**, 014057, doi:10.1103/PhysRevApplied.16.014057 (2021).

53  Yang, C. H., Lim, W. H., Zwanenburg, F. A. & Dzurak, A. S. Dynamically controlled charge sensing of a few-electron silicon quantum dot. *AIP Advances* **1**, 042111, doi:10.1063/1.3654496 (2011).

54  McJunkin, T. *Heterostructure Modifications, Fabrication Improvements, and Measurement Automation of Si/SiGe Quantum Dots for Quantum Computation.* PhD thesis, University of Wisconsin-Madison, (2021).

# Extended Data

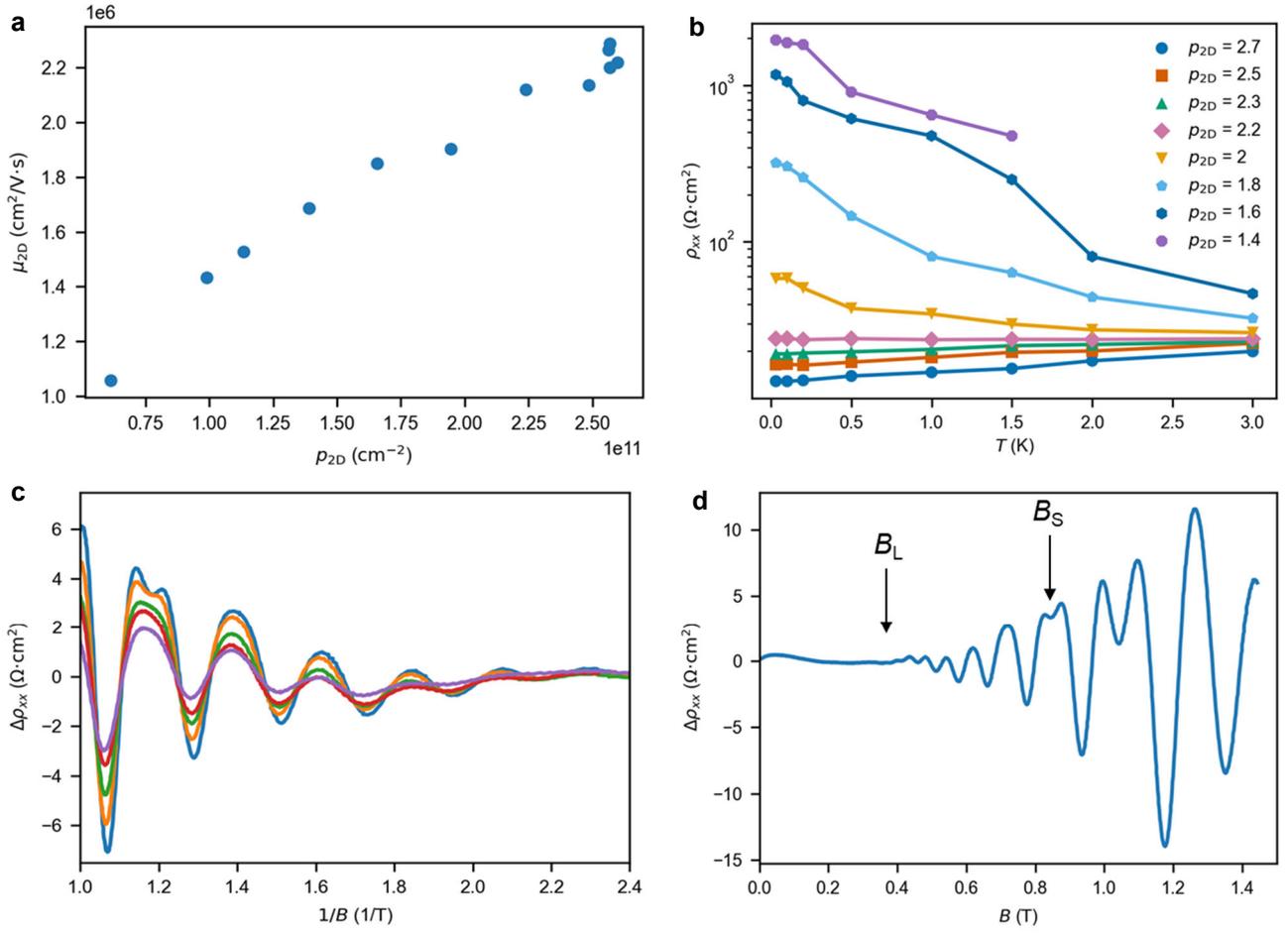

**Extended Data Fig. 1. Magnetotransport characterization of Hall-bar devices within a same Ge/SiGe heterostructure wafer at *T*=1.7 K.** **a,** Density ($p_{2D}$) dependence of mobility ($\mu_{2D}$). **b,** Temperature dependence of $\rho_{xx}$ at different $p_{2D}$ (with a unit of $10^{11}$cm$^{-2}$). A signature of MIT is observed, implying a strong hole-hole interaction in our 2DHG. **c,** Temperature dependence of the Shubnikov-de Haas oscillation $\Delta\rho_{xx}$ from $T$ = 1.7 K to 2.8K, after background subtractions. We obtain an effective mass $m^*$ = 0.09 $m_0$. **d,** g factor extraction from a Shubnikov-de Haas oscillation, after background subtractions. $B_L$ is the magnetic field at which the oscillation is visible and $B_S$ the Zeeman splitting are resolved. We get $g$ = 9.65.

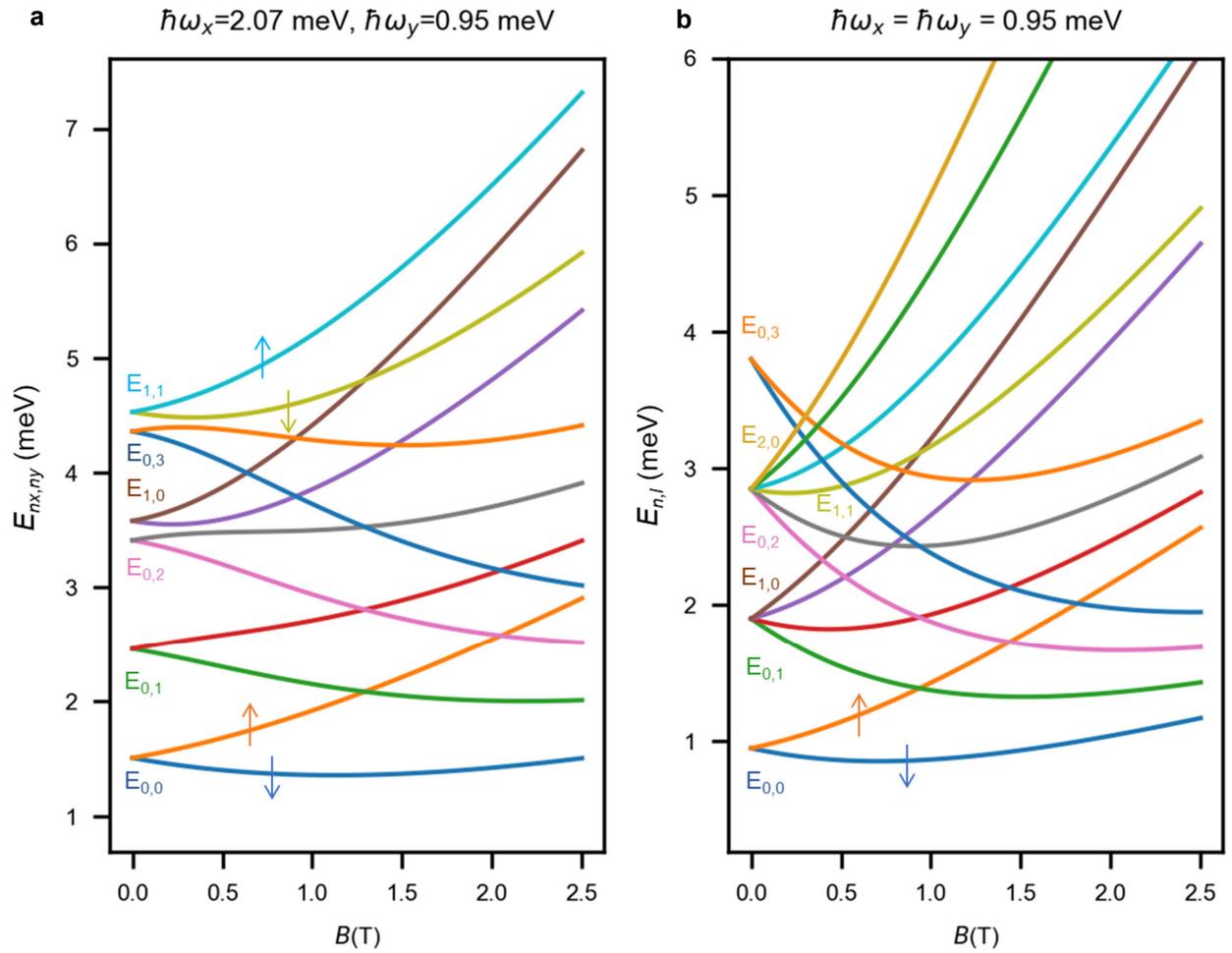

**Extended Data Fig. 2. Fock-Darwin eigenenergy for germanium quantum dot. a,** Simulation of magnetic field dependence of Fock-Darwin eigenenergy with an anisotropic parabolic confinement $\hbar\omega_x = 2.07$ meV and $\hbar\omega_y = 0.95$ meV (values taken from Extended Data. Fig. 3). We use $g = 9.65$ and $m^* = 0.09 m_0$ from magnetotransport characterization (see Extended Data Fig. 1). Arrows signify the spin direction of hole. The simulation for electrochemical potential in Fig. 2a is further obtained by considering a charging energy (giving rise to an offset for each hole occupancy) and a typical exchange interaction of 0.7 meV (pushing the avoid-cross (singlet to triplet transition for $N = 2$) from 1.3 T to 0.4 T to account for the experimental result in Fig. 2b). **b,** Simulation for Fock-Darwin eigenenergy with a circular parabolic confinement $\hbar\omega_x = \hbar\omega_y = 0.95$ meV. A negative-positive slope reversal before the first avoid-cross for $E_{0,1,\uparrow}$ (red) is a main difference for $\mu$ ($N = 4$) between circular and elliptical confinement.

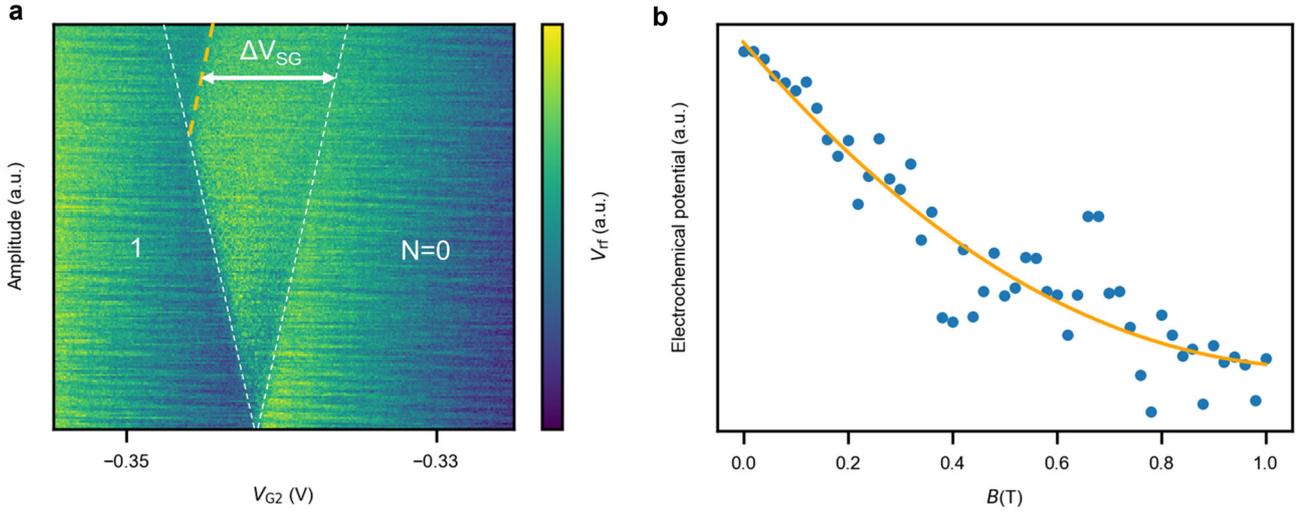

**Extended Data Fig. 3. Estimation of the strength of confinement potential for $d^* = 102$ nm. a,** Pulsed-gate spectroscopy of the G2 quantum dot for $d^* = 102$ nm. We apply a 1 MHz square wave through bias tee to the gate G2. An orange dashed line represents the first-excited state and the right white dashed line the ground state. We thus extract the first-excited energy $E_{orb} = \alpha \Delta V_{SG} = 1$ meV. **b,** Fit result for the spectrum of the first hole in Fig. 2b. The Fock-Darwin model with elliptical deformation is used. We take $g = 9.65$ and $m^* = 0.09 m_0$ from magnetotransport characterization (see Extended Data Fig. 1) and $\alpha = 0.115$ as fixed parameters. The result yields $\hbar \omega_x = 2.07$ meV and $\hbar \omega_y = 0.95$ meV. The energy (0.95 meV) of the first excited state ($n_x = 0$, $n_y = 1$) is in accordance with the pulse-gate measurement in **a**.

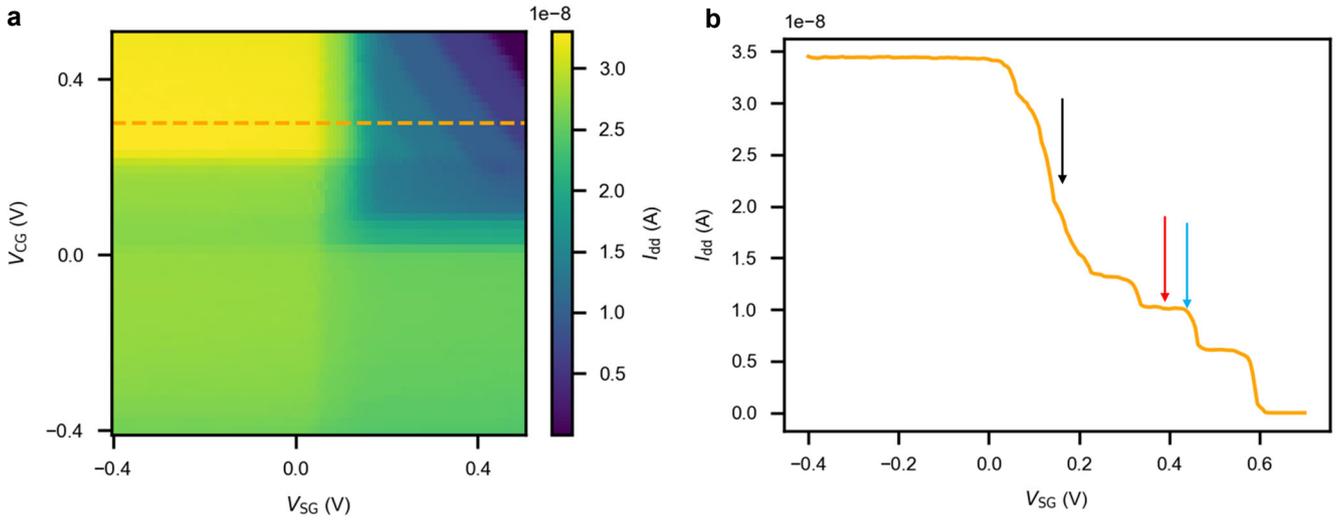

**Extended Data Fig. 4. Tuning method for dot size. a,** Color plot of the current through the double dot, as a function of the gate voltage of CG and SG. Another lock-in amplifier connected to the left and right bottom ohmic contacts (not shown in the Fig. 1b) is used to measure the current through the double dot at the coarse tuning period. Before we tune into the quantum dot regime, we first step $V_{CG}$ and sweep $V_{SG}$ to confine a one-dimensional channel after the device is pinch on. In the top right regime, there are a series of stripes, similar to the quantum point contact (QPC) steps. We can vary the confinement strength for the x-axis by choose the different voltage of SG. **b,** 1D-plot of the current from the orange dashed line in **a**. The black, red and blue arrows indicate the $V_{SG}$ in Fig. 2c ($d^* = 133$ nm), Fig. 2d ($d^* = 168$ nm) and Fig. 2b ($d^* = 102$ nm), respectively. The relatively small value of $V_{SG}$ applied in Fig. 2c effectively decreases the confinement strength for the x-axis, resulting the shape of quantum dot more circular.

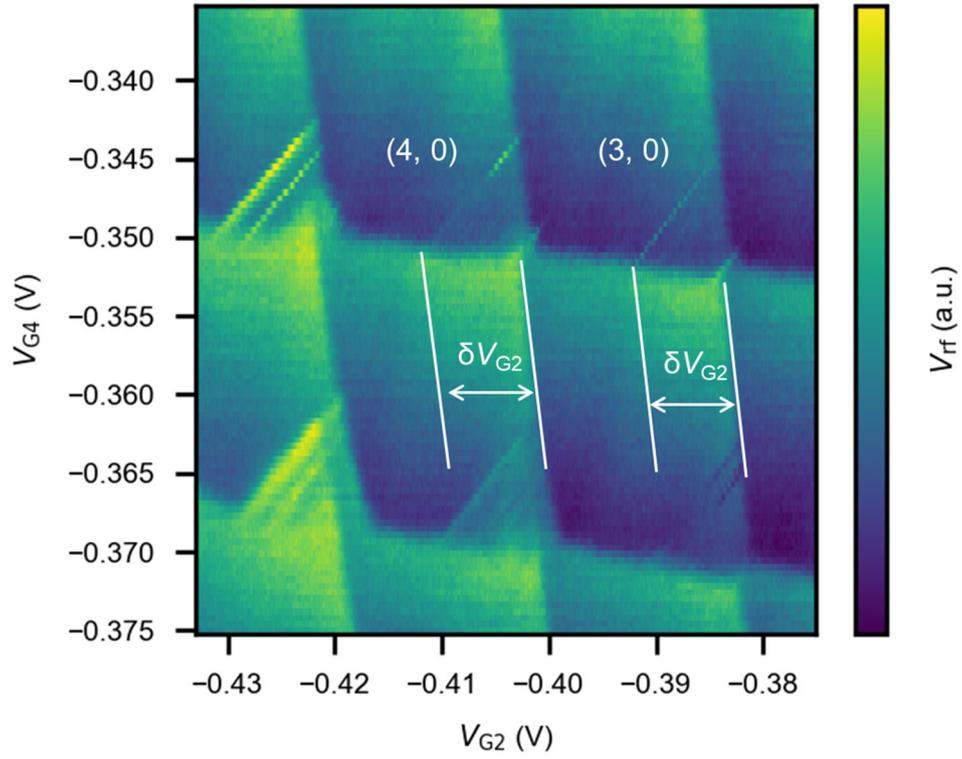

**Extended Data Fig. 5. Lever arm extraction.** We apply a bias of 1 mV to the source of double dot with drain grounded. Bias triangles for lower occupancies are unavailable due to small tunneling rate. We extract $\alpha = V_{SD}/\delta V_{G2} = 0.115$ eV/V for both (4, 0) and (3, 0) regions. Subsequently, we are confident that this value remains nearly unchanged for lower occupancies.